\documentclass[prl,twocolumn,showpacs,preprintnumbers,amsmath,amssymb,floatfix]{revtex4}

\usepackage[dvips]{graphicx}
\usepackage{dcolumn}
\usepackage{bm}

\begin{document}

\title{
The Role of the Exchange-Correlation Potential in {\it ab initio} \\
Electron Transport Calculations
}

\author{ San-Huang Ke,$^{1}$ Harold U. Baranger,$^{2}$ and Weitao Yang,$^{1}$}

\affiliation{
     $^{\rm 1}$Department of Chemistry, Duke University, Durham, NC 27708-0354 \\
     $^{\rm 2}$Department of Physics, Duke University, Durham, NC 27708-0305
}

\date{March 5, 2007; published as J. Chem.\,Phys.\,126, 201012 (2007)}

\begin{abstract}
The effect of the exchange-correlation potential in {\it ab initio} electron transport calculations is investigated by constructing optimized effective potentials (OEP) using different energy functionals or the electron density from second-order perturbation theory. We calculate electron transmission through two atomic chain systems, one with charge transfer and one without. Dramatic effects are caused by two factors: changes in the energy gap and the self-interaction error. The error in conductance caused by the former is about one order of magnitude while that caused by the latter ranges from several times to two orders of magnitude, depending on the coupling strength and charge transfer. The implications for accurate quantum transport calculations are discussed. 
\end{abstract}

\pacs{73.40.Cg, 72.10.-d, 85.65.+h}
\maketitle


The calculation of electron transport through single molecules directly from quantum mechanics is currently being intensively investigated for both fundamental physics and  applications in molecular electronics \cite{Heath0343}. In such a calculation, properties of the particular molecule must be incorporated into an accurate transport model. A frequently used theoretical approach is the single-particle Green function (GF) method \cite{Datta95} combined with a density functional theory (DFT) \cite{Parr89} electronic structure calculation. In this approach \cite{Taylor01245407,Damle01201403,Xue02151,Brandbyge02165401,Ke04085410}, the atomic structure of the entire lead-molecule-lead system is taken into account explicitly \cite{Brandbyge02165401,Ke04085410}. Despite its advantages and high efficiency for large systems, several aspects of this approach remain problematic \cite{Evers235411,Burke146803,Sai186810,Toher05146402,Koentopp121403,Muralidharan155410}. Here we address one aspect: we show that an improved description of electron-electron exchange and correlation within Kohn-Sham DFT dramatically changes the predicted conductance.

In the standard GF+DFT approach, all electron-electron interaction effects are incorporated through the self-consistent DFT calculation, while the transmission calculation is simply single-particle. Consequently, as emphasized by others \cite{Sai186810,Koentopp121403}, exchange-correlation corrections to the expression for the current are neglected. But even before considering those corrections, self-interaction error (SIE) is a potentially serious problem within the GF+DFT method itself \cite{Parr89,Toher05146402,Muralidharan155410}: it leads to an overly extended charge distribution and, therefore, inaccurate molecule-lead charge transfer, especially for weakly coupled systems. In {\it ab initio} transport calculations, SIE will directly affect the position of the chemical potential in the molecular HOMO-LUMO gap (``gap" for short), as well as the broadening of the HOMO and LUMO orbitals, possibly producing large errors in the conductance \cite{Toher05146402}. It is thus critical to improve the {\it xc} potential so as to eliminate the effects of SIE. Another well-known problem with DFT is that the predicted gap is too small, often leading to a significant overestimation of the conductance.
The solution to this problem relies on a quasiparticle calculation or the construction of a SIE-free functional with a nonlocal $xc$ potential. In this paper, we focus on eliminating the SIE and revealing the magnitude of the errors caused by the two problems.

One way to eliminate SIE is to use Hartree-Fock (HF) theory: the exact treatment of exchange eliminates SIE. However, because HF involves a single determinant and lacks dynamical screening, the LUMO orbital is not physically meaningful and the gap is too large for extended systems and large molecules. Hybrid functionals, like B3LYP \cite{Becke935648,Lee785}, are a possible compromise: these mix the HF exchange potential with the local effective potential obtained from the local density approximation (LDA) or generalized gradient approximation. Although B3LYP is a significant improvement over LDA for almost all molecular systems, SIE still remains \cite{Zhang982604}.

The optimized effective potential (OEP) approach is a direction for improving DFT calculations \cite{Yang143002}, in which the (local) effective potential is expressed as an implicit density functional in terms of the Kohn-Sham orbitals. OEP enables one to construct a local $xc$ potential from any energy functional, such as the HF exact exchange (EXX) or B3LYP functionals, or from an electron density obtained from a more accurate theory \cite{Wu2498}, such as second-order many-body perturbation theory (MP2). Most OEP calculations to date use the HF energy functional (EXX-OEP). This simplest exchange-only OEP approach improves systematically the electronic structure of various semiconductors \cite{Stadele10031}: the band gaps are significantly improved over those of both LDA and HF, although the underlying reason is still open \cite{Sharma05136402,Gruning06154108}.

\begin{figure}[t]
\includegraphics[angle= 0,width=6.0cm]{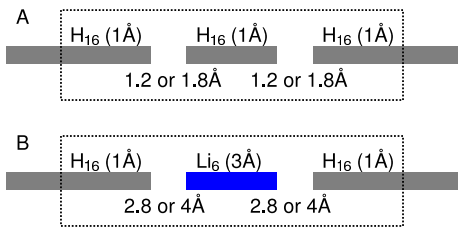}
\caption{
Schematic view of the atomic chain systems calculated. The extended molecule is
indicated by the dashed frame. The atomic separations in the lead, molecule,
and between them are indicated. Note that there is no charge transfer in system
A while there possibly is in system B.
}
\label{fig_str}
\end{figure}

In this paper, we implement the OEP approach in DFT-based {\it ab initio} transport calculations and investigate, for the first time, the effect of different {\it xc} potentials---LDA, HF, EXX-OEP, B3LYP, B3LYP-OEP, and MP2-OEP. Our purpose in using OEP is to construct a local {\it xc} potential which is SIE free (EXX-OEP and MP2-OEP), allowing us to assess the significance of the SIE and the energy gap in transport. We calculate electron transmission through two simple systems consisting of atomic chains (Fig.\,\ref{fig_str}), one in which molecule-lead charge transfer does not occur (system A) and one (system B) which does show such transfer.

The infinite open system is divided into three parts: left and right leads, and device region. The latter includes large parts of the leads to accommodate the molecule-lead interaction and so achieve good convergence. The electronic structure calculations are carried out in a cluster geometry \cite{NWChem}, both for the traditional {\it xc} potentials -- LDA, HF exchange, and B3LYP -- and for MP2. In all these calculations, the wave function is expanded in the Gaussian basis set 6-311G**. For system A, the cluster is H$_{128}$; for B, it is H$_{56}$--Li$_6$--H$_{56}$. Both are much larger than the device region (Fig.\,\ref{fig_str}) to ensure that the middle part of the left and right H-chain has the proper bulk lead properties. Because our focus is the device region, the electronic structure of the left and right leads is described by LDA in all the calculations.

The OEP approach using HF and B3LYP energy functionals has been implemented in the efficient way described in Ref.\,\onlinecite{Yang143002}. For MP2-OEP, we find the OEP from the electron density given by an MP2 calculation \cite{Wu2498}. In our calculations, the OEP [$v_{\rm eff}({\mathbf r})$] is expanded in the 6-311G** Gaussian basis set. (Test calculations using cc-pvdz and cc-pvtz show only minor difference.)

The self-consistent $v_{\rm eff}(\mathbf{r})$ determines the optimized Hamiltonian $\mathbf{H}_{D}$ of the device region. In terms of the basis states $\{\phi_i({\mathbf r})\}$ with overlap matrix ${\mathbf S}$, the retarded Green function of the device region is $\mathbf{G}_{D}(E)=\big[ E^{+}\mathbf{S}_{D} - \mathbf{H}_{D}-{\Sigma}_{L}(E) -{\Sigma}_{R}(E) \big]^{-1}$, where ${\Sigma}_{L,R}(E)$ is the self-energy for the semi-infinite left or right lead. The electron transmission coefficient at any energy, $T(E)$, is calculated from the Green function, and the conductance, $G$, then follows from a Landauer-type relation \cite{Datta95,Ke04085410}.


\begin{figure}[t]
\includegraphics[angle=-0,width=6.5cm]{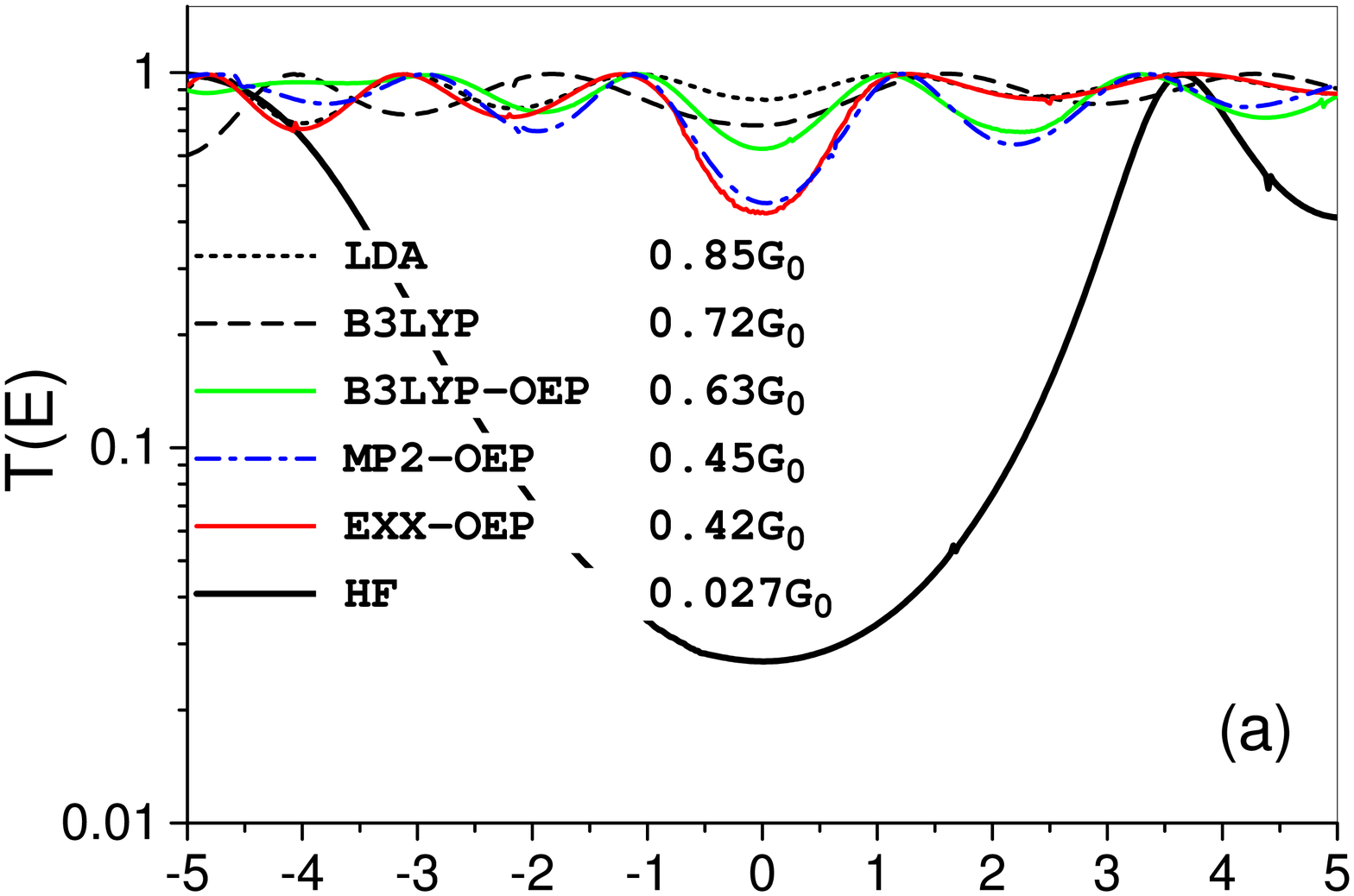} \\
\includegraphics[angle=-0,width=6.5cm]{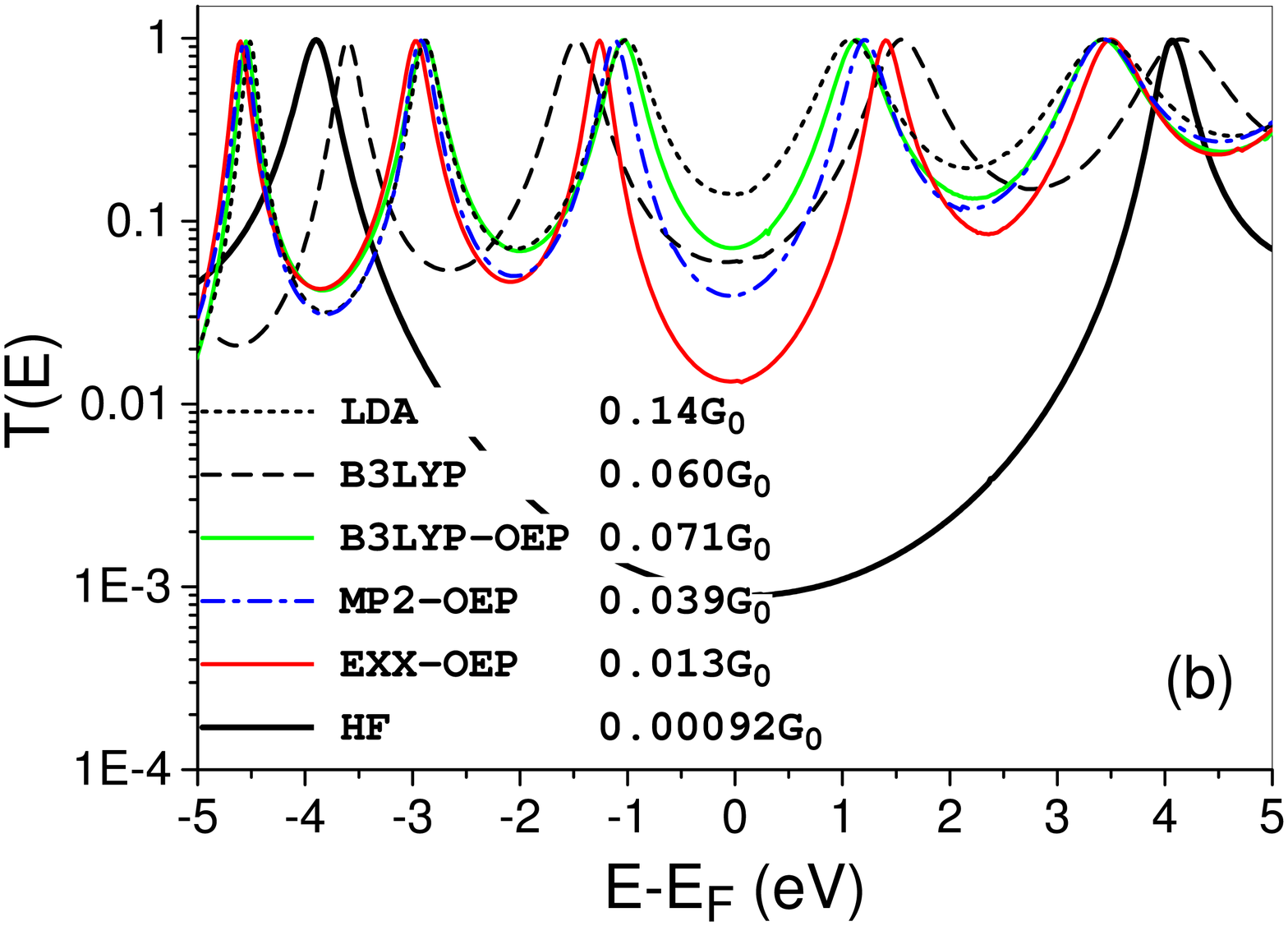}
\caption{
(color online) Transmission through system A with (a) 1.2\,{\AA} and (b)
1.8\,{\AA} molecule-lead separation. Results for six functionals with different {\it xc} potentials are shown, and the equilibrium conductance is listed in the legend.
}
\label{fig_a}
\end{figure}

The transmission through system A is shown in Fig.\,\ref{fig_a} for both a typical short molecule-lead separation (1.2\,{\AA}) and a longer separation (1.8\,{\AA}) corresponding to weaker coupling. Because there is no molecule-lead charge transfer, the Fermi energy is right at the middle of the gap. For both values of the molecule-lead separation, the results divide into two classes: the HF result and the others. The very small HF conductance is related to the gap being much larger.

Let us first look at the SIE issue by dividing the results in the second class into three groups: with SIE (LDA), partially with SIE (B3LYP and B3LYP-OEP), and without SIE (EXX-OEP and MP2-OEP), denoted hereafter by groups I, II, and III, respectively. From Fig. 2, increased SIE leads to increased conductance: $G_{\rm group\,I} > G_{\rm group\,II} > G_{\rm group\,III}$. This trend follows from noting that SIE overly broadens the HOMO and LUMO states. It is clear in Fig. 2 that this trend is dominated by the SIE regardless of the gap behavior; for example, the B3LYP gap is noticeably larger than the MP2-OEP gap but the MP2-OEP conductance is much smaller than the B3LYP one. This indicates that within DFT the SIE issue, rather than the gap issue, is more important in determining the conductance. Thus, EXX-OEP and MP2-OEP should be a significant improvement over the other $xc$ potentials for DFT-based 
transport calculation.

For smaller separation (stronger coupling), Fig.\,\ref{fig_a}(a), all the {\it xc} potentials give a similar conductance -- the maximum difference (between LDA and EXX-OEP) is only about a factor of $2$. This is understandable because SIE mainly affects systems containing localized electrons. 
As the coupling becomes weaker, Fig.\,\ref{fig_a}(b), the peaks in $T(E)$
become sharper and the conductance decreases. Note that the spread in conductance values becomes larger: now the maximum difference (between LDA and EXX-OEP) is about a factor of $10$. 

To show the effect from the gap, we first examine the real transport gap of the
H$_{16}$ molecule by calculating the ionization potential ($I$) and electron
affinity ($A$) using the delta self-consistent field method ($\Delta$SCF) and
the outer valence Green function method (OVGF) \cite{Ortiz886348}. The result
for $I\!-\!A$ is $\Delta$SCF(HF, LDA, B3LYP)= 4.7eV, 6.0eV, 5.9eV, and OVGF=
6.6eV. Note that all the results are substantially smaller than the 8eV HF gap,
indicating that it is too large. In particular, the large difference between
$\Delta$SCF(HF) and HF shows that HF does not give a good description for this
long-chain molecule because of the lack of screening/correlation, despite the fact that it works well for very small molecules (the screening is very weak there, see the database at http://srdata.nist.gov/cccbdb). On the other hand, $I\!-\!A$ is significantly larger than all the DFT gaps (2 $\sim$ 3 eV) in the second class, indicating that they are too small (EXX-OEP does not work well here). A rough estimation of the effect of the gap is the difference in conductance between EXX-OEP and HF, both of which lack correlation and are SIE free. In Fig. 2, this difference is about one order of magnitude for both strong and weak coupling; because the HF gap is too large, this rough estimation is probably an overestimate.

\begin{figure}[t]
\includegraphics[angle=  0,width=6.0cm]{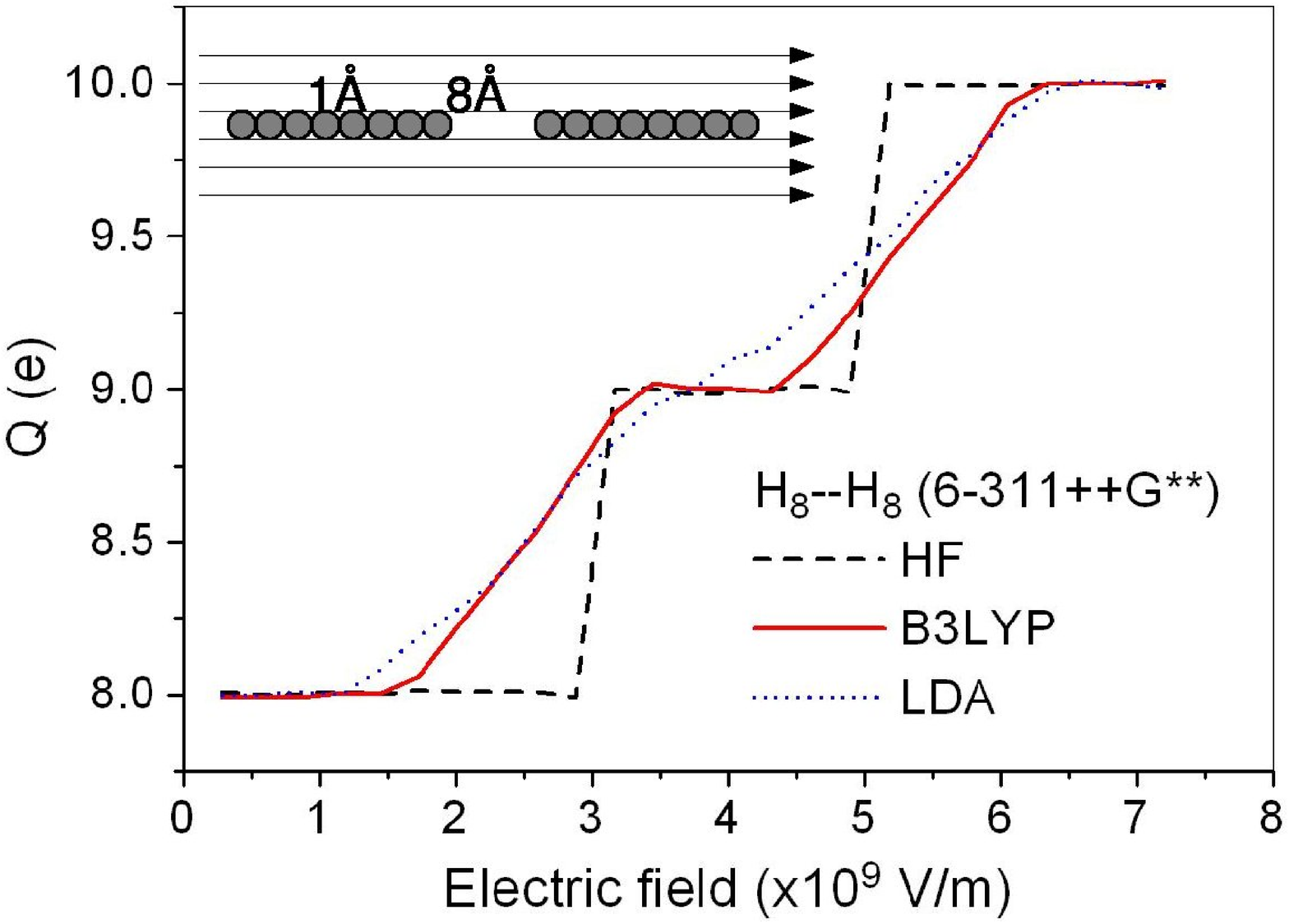}
\caption{
(color online) Electric field induced electron transfer (change in $Q$) between 
two H$_8$ clusters separated by 8\,{\AA} (as shown in the inset). For such a
large separation, the electron transfer should be an integer. For HF, this is
the case; however, LDA and B3LYP calculations show a substantial SIE. 
}
\label{fig_q}
\end{figure}

So far we have discussed the SIE and gap issues for the system without molecule-lead charge transfer, where SIE causes overly broadened HOMO and LUMO states. For systems with charge transfer, SIE is a more significant problem because it may lead to too much charge transfer, particularly in weakly coupled systems. To directly demonstrate this, we calculate, by using Mulliken population analysis, the charge transfer between two weakly coupled H atomic chains induced by a strong electric field (see Fig.\ref{fig_q}). Each chain contains 8 H atoms separated by 1\,{\AA}, and the separation between the two chains is 8\,{\AA}. Because of the very large separation, the physical electron transfer must be an integer. In HF, the electron transfer is indeed always an integer, showing that it is SIE free. For LDA, the result is almost linear in the electric field (full SIE), while B3LYP significantly improves upon LDA but is still not accurate (partial SIE). 

\begin{figure}[t]
\includegraphics[angle=-0,width=6.5cm]{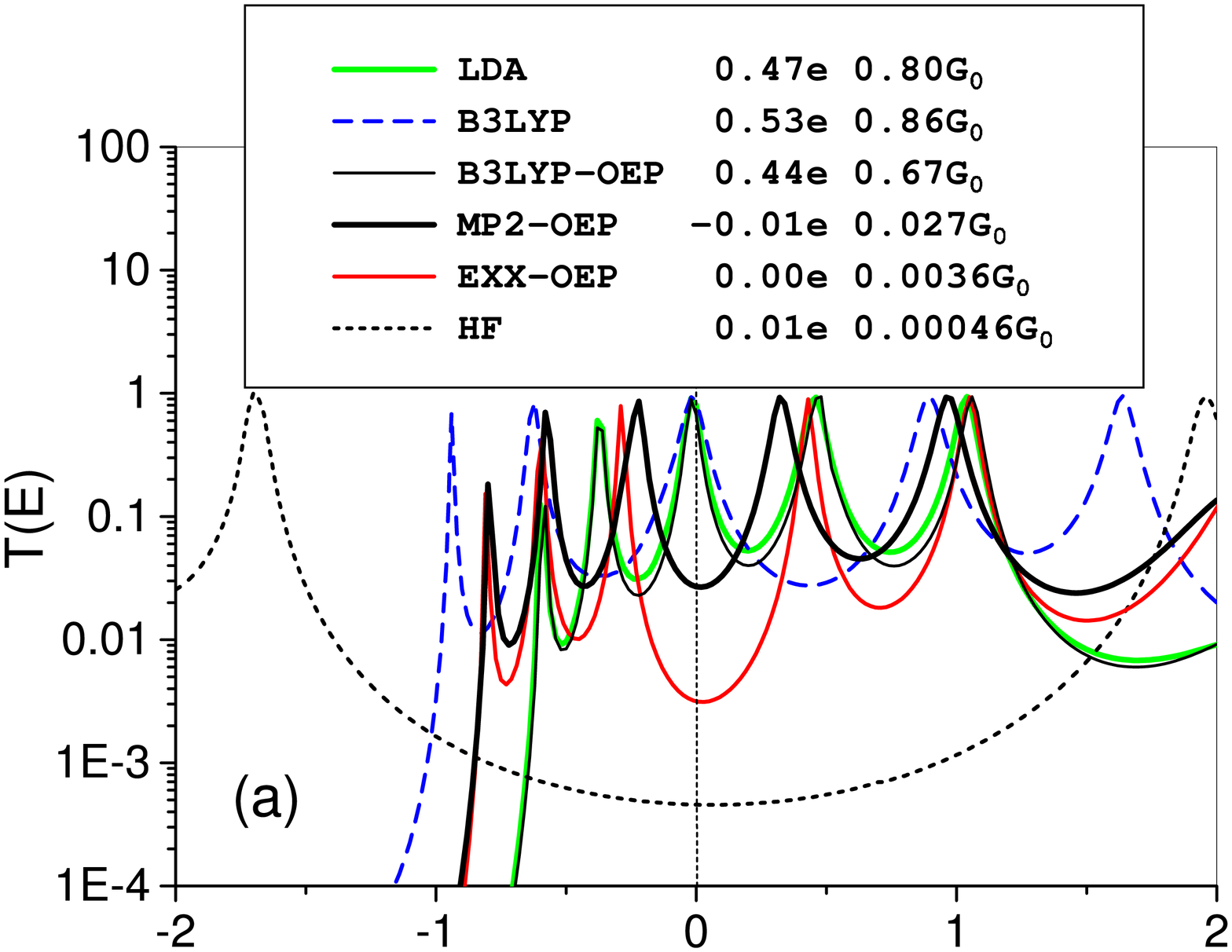} 
\includegraphics[angle=-0,width=6.5cm]{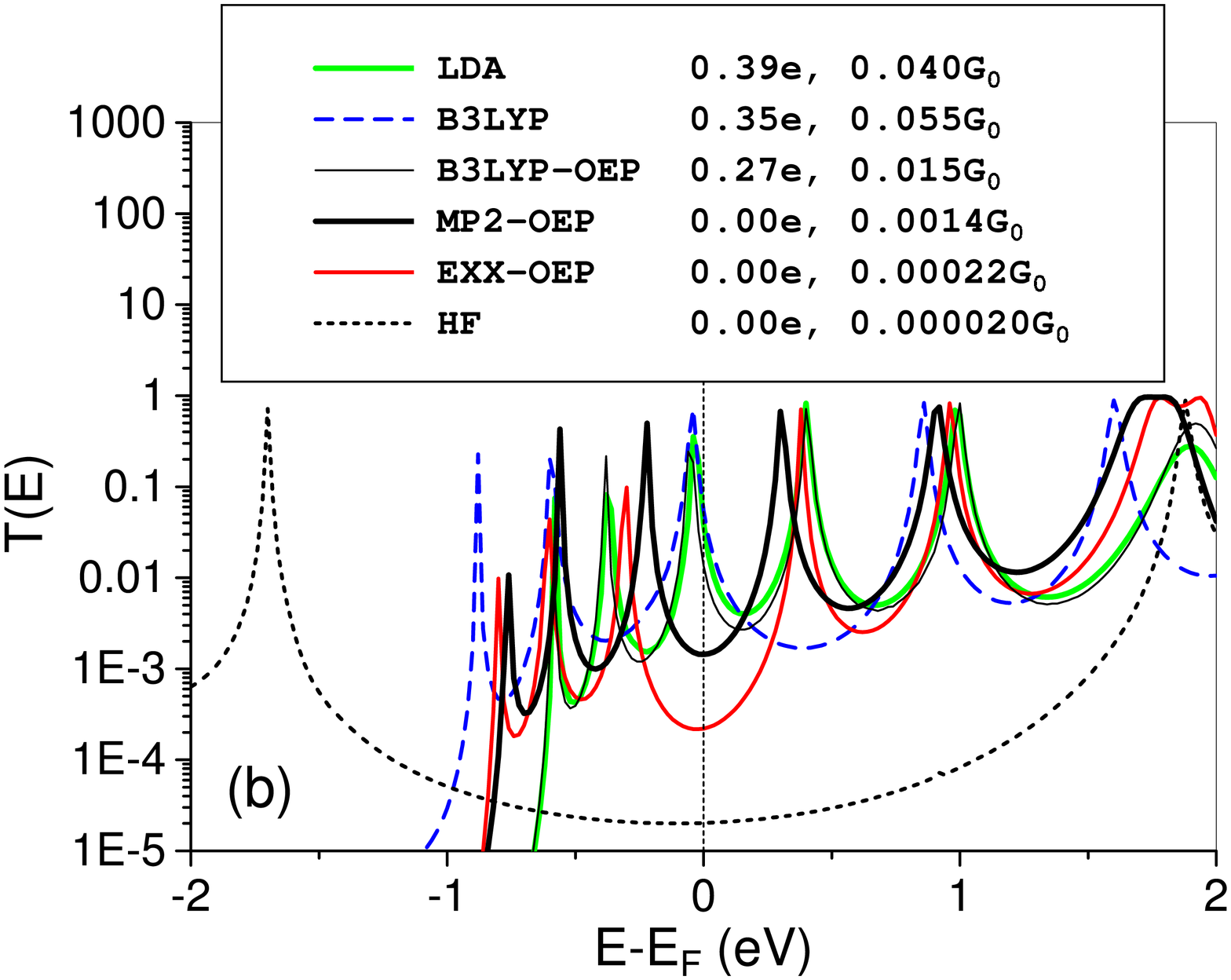}
\caption{
(color online) Transmission through system B with (a) 2.8\,{\AA} and (b)
4.0\,{\AA} molecule-lead separation for six energy functionals with different
{\it xc} potentials. In the legend, the equilibrium conductance and the charge
transfer from the Li cluster to the H-chain are listed. The SIE-infected
functionals place the chemical potential near a molecular resonance, while the
SIE-free functionals place it near the middle of the gap. 
}
\label{fig_b}
\end{figure}

The transmission through system B, in which there may be substantial charge
transfer, is shown in Fig.\,\ref{fig_b} for two values of the molecule-lead
separation. Molecule-lead charge transfer determines the position
of the chemical potential (fixed in the lead) in the molecular gap, and
therefore the resulting conductance. The charge transfer and conductance are
listed in the figure. Note the strikingly different behavior of the two groups
of functionals: for functionals with SIE (LDA, B3LYP, and B3LYP-OEP), the
chemical potential enters the HOMO resonance because the charge
transfer is large, while for functionals without SIE (HF, EXX-OEP, and MP2-OEP), the
chemical potential is at the middle of the gap because the charge
transfer is near zero. Consequently, the conductance given by these two groups of
functionals are \textit{very} different, up to three orders of magnitude. 

For the smaller separation, 2.8\,{\AA} in Fig.\,\ref{fig_b}(a), the functionals with SIE give a charge transfer of about $0.5\,e$, and the resulting conductance is around $0.8\,G_0$. When the separation is increased to 4.0\,{\AA} [panel (b)], the peaks become sharper, and the conductance in all cases decreases by more than an order of magnitude. In contrast, the charge transfer resulting from the SIE functionals decreases only slightly, showing clearly that it is an artifact of SIE. Despite the quantitative differences between the stronger and weaker coupling, the broad features in the two cases are the same: the biggest step in conductance (a factor of $\sim$30) is between the SIE functionals and MP2-OEP followed by two smaller decreases, first from MP2-OEP to EXX-OEP and then further to HF, each by about a factor of 10. Here the effect from the gap is also about one order of magnitude, from comparing EXX-OEP and HF as for system A.

While it is clear, in terms of SIE, that the EXX-OEP and MP2-OEP calculations improve significantly the standard GF+DFT calculation, it is not obvious which one of the SIE-free functionals -- HF, EXX-OEP, or MP2-OEP -- gives a conductance closest to the truth. MP2-OEP provides a near-exact local effective potential for Kohn-Sham DFT, but its finite $xc$ potential discontinuity is not included in the gap. As a result, its gap is too small. EXX-OEP gives a slightly larger gap which, however, is still too small, and correlation is absent. HF, on the other hand, yields too large a gap, and correlation is also absent. Therefore, in terms of transport, EXX-OEP and MP2-OEP probably overestimates the conductance while HF probably underestimates it. The error seems to be about a factor of 10.

Finally, we relate the present calculation to the more rigorous time-dependent DFT formalism (TDDFT) \cite{Stefanucci0414,Stefanucci0607333}. In principle, unlike DFT, TDDFT can treat the electronic structure of excited states \cite{Runge84997}. A Landauer-like form for the steady-state current can be derived from TDDFT \cite{Stefanucci0607333}: in the linear-response regime (zero bias), the current is a Kohn-Sham term plus a correction from dynamical $xc$ effects. The effective potential in our calculation can be regarded as the long time limit of that in the Kohn-Sham term, which is the major part of the current. The missing dynamical $xc$ effect is an open issue studied in \cite{Sai186810,Koentopp121403}. Our results on the effects of the $xc$ potential are helpful for improving TDDFT calculations within the adiabatic approximation.

In summary, by implementing the OEP approach in an {\it ab initio} transport calculation, we have systematically investigated the effect of different local and nonlocal $xc$ potentials.
Dramatic effects, up to orders of magnitude, originate from two factors -- the
SIE and the energy gap. The former will dominate for systems with charge
transfer and can be eliminated by using a SIE-free OEP potential, while the
latter is difficult to treat within DFT and leads to a typical overestimation
of the conductance by about a factor of 10. Possible solutions are either to
perform transport calculations using quasiparticle states, like those in GW
approximation, or to develop SIE-free energy functionals without 
the discontinuity problem.

%

We thank Aron Cohen, Kieron Burke, and Qin Wu for valuable conversations. This work was supported in part by the NSF (DMR-0506953). 


\end{document}